\documentclass[11pt]{article}
\usepackage{newclude}
\usepackage{amssymb,amsmath}
\usepackage{color}
\usepackage{graphicx}
\usepackage[numbers]{natbib}
\usepackage{subcaption} 

\usepackage{hyperref}

\usepackage{fancyvrb}

\usepackage{pdfpages}
\usepackage{listings}
\newsavebox{\fmbox}

\usepackage{enumitem}

\usepackage{float}

\usepackage[left=4cm, right=4cm, top=2cm]{geometry}

\usepackage[table]{xcolor} 
\usepackage{multirow}

\usepackage{algorithm, setspace}
\usepackage[noend]{algpseudocode}
\DefineVerbatimEnvironment{Stancode}{Verbatim}{fontsize=\small,xleftmargin=2em}  

\usepackage{tikz}
\usetikzlibrary{positioning}

\usepackage{epigraph}

\makeatletter 
\renewcommand\@biblabel[1]{#1.} 
\makeatother

\usepackage{listings}
\definecolor{codegreen}{rgb}{0,0.6,0}
\definecolor{codegray}{rgb}{0.5,0.5,0.5}
\definecolor{codepurple}{rgb}{0.58,0,0.82}
\definecolor{backcolour}{rgb}{0.99,0.99,0.97}
\lstdefinestyle{stan}{
	literate={~}{$\sim$}{1},
	backgroundcolor=\color{backcolour}, 
	comment=[l]{//},
	commentstyle=\color{codegreen},
	keywords = {int, real, vector, matrix, data, model, parameters, transformed, target, generated, quantities, for, in},
	keywordstyle=\color{magenta},
	numberstyle=\tiny\color{codegray},
	stringstyle=\color{codepurple},
	emph={%
		normal, normal_rng,
		pmx_solve_twocpt,
		cmdstan_model,
		exp, log
	},
	emphstyle=\color{codepurple},%
	basicstyle={\ttfamily},
	breaklines=true,                 
	keepspaces=true,                 
	showspaces=false,                
}
\title{Flexible and efficient Bayesian pharmacometrics modeling using Stan and Torsten, Part I}
\author{Charles C. Margossian$^1$, Yi Zhang$^2$, and William R. Gillespie$^2$}
\date{$^1$Department of Statistics, Columbia University (formally Metrum Research Group, LLC)\\
  $^2$Metrum Research Group, LLC\\
The authors declared no competing interests for this work.}

\begin{document}

\maketitle

\begin{abstract}
Stan is an open-source probabilistic programing language, primarily
designed to do Bayesian data analysis.  Its main inference algorithm
is an adaptive Hamiltonian Monte Carlo sampler, supported by state of
the art gradient computation.  Stan's strengths include efficient
computation, an expressive language which offers a great deal of
flexibility, and numerous diagnostics that allow modelers to check
whether the inference is reliable.  Torsten extends Stan with a suite
of functions that facilitate the specification of pharmacokinetic and
pharmacodynamic models, and makes it straightforward to specify a
clinical event schedule.  Part I of this tutorial demonstrates how to
build, fit, and criticize standard pharmacokinetic and pharmacodynamic
models using Stan and Torsten.
\end{abstract} 

\section{Introduction}

Bayesian inference offers a principled approach to learn about unknown variables from data using a probabilistic analysis.
The conclusions we draw are based on the posterior distribution which, in all but the simplest cases, is intractable.
We can however probe the posterior using a host of techniques such as Markov Chains Monte Carlo sampling and approximate Bayesian computation.
Writing these algorithms is a tedious and error prone endeavor but fortunately modelers can often rely on existing software with efficient implementations.

In the field of pharmacometrics, statistical software such as
NONMEM\textsuperscript{\textregistered} \cite{Beal_undated-qu},
Monolix\textsuperscript{\textregistered} \cite{monolix2021},
{and the R package nlmixr \cite{fidler_nonlinear_2019}}
support many routines to specify and analyze pharmacokinetic and
pharmacodynamic population models.
There also exist more general probabilistic programing languages such as BUGS \cite{lunn2009bugs} and more recently Stan \cite{Carpenter:2017}, to only name a few examples.
This tutorial focuses on Stan.
Stan supports a rich library of probability densities, mathematical functions including matrix operations and numerical solvers for differential equations.
These features make for an expressive and flexible language, however writing common pharmacometrics models can be tedious.
Torsten extends Stan by providing a suite of functions to facilitate the specification of pharmacometrics models.
These functions make it straightforward to model the event schedule of a clinical trial and parallelize computation across patients for population models.

{
This tutorial reviews key elements of a Bayesian modeling workflow in Stan,
including model implementation,
inference using Markov chains Monte Carlo (MCMC),
and diagnostics to assess the quality of our inference and modeling.
We assume the reader is familiar with compartment models in pharmacokinetics and pharmacodynamics and has experience with data that describe a clinical event schedule.
Since Torsten follows the input conventions in NMTRAN\textsuperscript{\textregistered}, experience with
NONMEM\textsuperscript{\textregistered} is helpful, though not essential.
Likewise, exposure to Bayesian statistics and inference algorithms is desirable, in particular an elementary understanding of MCMC.

We introduce programming in Stan and Torsten with the assumption that the reader is familiar with R. }


\subsection{Why Stan?}

We believe that Stan, coupled with Torsten, can be an important addition to the pharmacometrician's toolkit, especially for Bayesian data analysis.

The most obvious strength of Stan is its flexibility: it is straightforward to specify priors, systems of ODEs, a broad range of measurement models, missing data models and complex hierarchies (i.e. population models).
{
Examples of how Stan's flexibility may be leveraged in pharmacometrics include:
\begin{itemize}
  \item Combining various sources of data and their corresponding measurement models into one large model, over which full Bayesian inference can be performed \cite[e.g.][]{Weber:2018}.
  In a similar vain, it is possible to build complex hierarchical structures, which allow us to simultaneously pool information across various groups, e.g. patients, trials, countries.
  We will study such an example in Part II of this tutorial.
  \item Using a sparsity inducing prior, such as a the \textit{Horseshoe} prior \cite{Carvalho+Polson+Scott:2010:HS, Piironen:2017}, to fit models with a high-dimensional covariate.
  This approach has, for example, been used in oncology \cite{Yin:2021}
  and is a promising avenue in pharmacogenetics \cite{Bertrand:2017}.
  \item Incorporating a non-parametric regression, such as a Gaussian process, to build a translational model for pediatric studies \cite[e.g.][]{Siivola:2021}.
\end{itemize}
Stan's expressive language plays a crucial part here, since more specialized software do not readily handle the relatively complex structures and priors the above examples require.
}

%

Secondly, Stan supports state of the art inference algorithms, most notably an adaptive \textit{Hamiltonian Monte Carlo} (HMC) sampler, a gradient-based Markov chains Monte Carlo (HMC) algorithm \citep{Betancourt:2018} based on the No U-Turn sampler (NUTS) \cite{Hoffman:2014}, automatic differentiation variational inference (ADVI) \cite{Kucukelbir:2017}, and penalized maximum likelihood estimators.
Stan's inference algorithms are supported by a modern automatic
differentiation library that efficiently generates the requisite derivatives \cite{Carpenter:2015}.
It is worth pointing out that algorithms such as NUTS and ADVI were first developed and implemented in Stan, before being widely adopted by the applied statistics and { modeling} communities.
As of the writing of this article, new inference algorithms continue
to be prototyped in Stan. Recent such examples include adjoint--differentiated Laplace approximations \cite{Margossian:2020}, cross-chain warmup \cite{Zhang:2020}, and path finding for improved chain initialization \cite{Zhang:2021}. 
{ Some of Stan's algorithms are now available in specialized pharmacometrics software.
NONMEM\textsuperscript{\textregistered} supports an HMC sampler,
although certain diagnostics required to assess the quality of HMC, notably for population models, are still missing.}

Stan indeed provides a rich set of diagnostics, including the detection of divergent transitions during HMC sampling \cite{Betancourt:2018}, and the improved computation of effective sample sizes and scale reduction factors, $\hat R$ \cite{Vehtari:2020}, as well as detailed warning messages based on these diagnostics.
The automated running of these diagnostics makes the platform more user-friendly and provides much guidance when troubleshooting our model and our inference.

Last but not least, both Stan and Torsten are open-source projects, meaning they are free and their source code can be examined and, if needed, scrutinized. 
The projects are under active development with new features being added regularly.


\subsection{Bayesian inference: notation, goals, and comments}

Given observed data $\mathcal D$ and latent variables $\theta$ from the
parameter space $\Theta$, a Bayesian model is defined by the joint distribution $p(\mathcal D, \theta)$.
The latent variables can include model parameters, missing data, and more.
In this tutorial, we are mostly concerned with estimating model parameters.

The joint distribution observes a convenient decomposition,
\begin{equation*}
  p(\mathcal D, \theta) = p(\theta) p(\mathcal D \mid \theta),
\end{equation*}
with $p(\theta)$ the \textit{prior} distribution and $p(\mathcal D \mid \theta)$ the \textit{likelihood}.
The prior encodes information about the parameters, usually based on scientific expertise or results from previous analysis.
The likelihood tells us how the data is distributed for a fixed parameter value and, per one interpretation, can be thought of as a ``story of how the data is generated'' \cite{Gelman:2013b}.
The Bayesian proposition is to base our inference on the \textit{posterior} distribution of the parameters, $p(\theta \mid \mathcal D)${, and more generally the posterior distribution of any derived quantity of interest, $p(f(\theta) \mid \mathcal D)$.}



For typical pharmacometric applications, the full joint posterior density of the model parameters is an unfathomable object which lives in a high dimensional space.
Usually we cannot even numerically evaluate the posterior density at any particular point!
Instead we must probe the posterior distribution and learn the characteristics that interest us the most.
In our experience, this often includes a measure of a central tendency and a quantification of uncertainty, for example the mean and the variance, or the median and the $5^\mathrm{th}$ and $95^\mathrm{th}$ quantiles { for any quantity of interest}.
For skewed or multimodal distributions, we may want a more refined analysis which looks at many quantiles.
What we compute are estimates of these quantities. 
Most Bayesian inference involves calculations based on marginal posterior distributions. That typically requires integration over a high number of dimensions---an integration that is rarely tractable by analytic or numerical quadrature.
One strategy is to generate \textit{approximate} samples from the posterior distribution and then use sample mean, sample variance, and sample quantiles as our estimators.

Bayes' rule teaches us that
\begin{equation*}
  p(\theta \mid \mathcal D) = \frac{p(\mathcal D, \theta)}{p(\mathcal D)} =  \frac{p(\mathcal D \mid \theta) p(\theta)}{p(\mathcal D)}.
\end{equation*}
Typically we can evaluate the joint density in the numerator but not the normalizing constant, $p(\mathcal D)$, in the denominator.
A useful method must therefore be able to generate samples { from the posterior $p(\theta \mid \mathcal D)$} using the \textit{unnormalized} posterior density, $p(\mathcal D, \theta)$.
{ Once we generate a sample $\theta$, we can apply a transformation $f$ to obtain a sample from $p(f(\theta) \mid \mathcal D)$.}

Many MCMC algorithms are designed to generate samples from an unnormalized density.
Starting at an initial point, these chains explore the parameter space $\Theta$, one iteration at a time, to produce the desired samples.
{
The first iterations of MCMC are used to find and explore the region in the parameter space where the posterior probability mass concentrates.
Only after this initial \textit{warmup phase}, do we begin the \textit{sampling phase}.
}
%
Hamiltonian Monte Carlo (HMC) is an MCMC method which uses the gradient to efficiently move across the parameter space \cite{Neal:2012, Betancourt:2018}.
Computationally, running HMC requires evaluating $\log p(\mathcal D, \theta)$ and $\nabla_\theta \log p(\mathcal D, \theta)$ many times across $\Theta$, i.e. for varying values of $\theta$ but fixed values of $\mathcal D$.
For this procedure to be well-defined, $\theta$ must be a continuous variable, else the requisite gradient does not exist.
Discrete parameters require a special treatment, which we will not discuss in this tutorial.

A Stan program specifies a method to evaluate $\log p(\mathcal D, \theta)$.
Thanks to automatic differentiation, this implicitly defines a procedure to compute $\nabla_\theta \log p(\mathcal D, \theta)$ \cite{Griewank:2008, Baydin:2018, Margossian:2019}.
Together, these two objects provide all the relevant information about our model to run HMC sampling and other gradient-based inference algorithms.

\subsection{Bayesian workflow}

Bayesian inference is only one step of a broader modeling process, which we might call the Bayesian workflow \cite{Betancourt:2018, Gabry:2019, Gelman:2020}.
Once we fit the model, we need to check the inference and if needed, fine tune our algorithm, or potentially change method.
And once we trust the inference, we naturally need to check the fitted model.
Our goal is to understand the shortcomings of our model and motivate useful revisions.
During the early stages of model development, this mostly comes down to troubleshooting our implementation and later this ``criticism'' step can lead to deeper insights.

All through the tutorial, we will demonstrate how Stan and Torsten can be used to check our inference and our fitted model.

\subsection{Setting up Stan and Torsten}

Detailed instructions on installing Stan and Torsten can be found on \url{https://github.com/metrumresearchgroup/Torsten}.
At its core, Stan is a C++ library but it can be interfaced with one of many scripting languages, including R, Python, and Julia.
{
Running Stan requires a modern C++ compiler such as g++ 8.1 provided by RTools 4.0 on Windows and the GNU-Make utility program on Mac or the Windows equivalent mingw32-make.
More details of setting up work environment can be found in \cite{cmdstan_guide:2021}.
}
We will use cmdStanR, which is a lightweight wrapper of Stan in R, and
in addition, the packages posterior \cite{Bukner:2020}, bayesplot \cite{Gabry:2021}, and loo \cite{Gabry:2020}.
We generate most of the figures in this paper using BayesPlot, though at times we trade convenience for flexibility and fall back to ggplot2 \cite{Wickham:2009}.

The R and Stan code for all the examples are available at \url{https://github.com/metrumresearchgroup/torsten_tutorial_1_supplementary}.

 \subsection{Resources} 
 Helpful reads include the \textit{Stan User Manual} \cite{Stan:2021} and the \textit{Torsten User Manual} \cite{Torsten:2021}. 
\textit{Statistical Rethinking} by McElreath (2020)\cite{mcelreath2020statistical} provides an excellent tutorial on Bayesian analysis that may be used for self-learning.
 A comprehensive textbook on Bayesian modeling is \textit{Bayesian Data Analysis} by Gelman et al (2013) \cite{Gelman:2013b}, with more recent insights on the Bayesian workflow provided by Gelman et al (2020) \cite{Gelman:2020}. 
 Betancourt (2018) \cite{Betancourt:2018} offers an accessible discussion on MCMC methods, with an emphasis on HMC.

\section{Two compartment model}  \label{sec:twoCpt}

As a starting example, we demonstrate the analysis of longitudinal plasma drug concentration data from a single individual using a linear two--compartment model with first order absorption.
The individual receives multiple doses at regular time intervals and
the plasma drug concentration is recorded over time.
Our goal is to estimate the posterior distribution of the parameters of the model describing the time course of the plasma drug concentrations in this individual.

\subsection{Pharmacokinetic model and clinical event schedule} \label{sec:twocpt}

Let us assume an individual receives a drug treatment of 1200 mg boluses q12h$\times$14 doses. Drug
concentrations are measured in plasma obtained from blood sampled at
0.083, 0.167, 0.25, 0.5, 0.75, 1, 1.5, 2, 3, 4, 6 and 8 hours
following the first, second and final dose.
In addition, we take measurements before each drug intake, as well as 12, 18, and 24 hours following the last dose.
We analyze that data using a two--compartment model with first
order absorption:
\begin{subequations}
\begin{eqnarray}
  \frac{\mathrm d u_\mathrm{gut}}{\mathrm d t} & = & - k_a u_\mathrm{gut} \\ 
  \frac{\mathrm d u_\mathrm{cent}}{\mathrm d t} & = & k_a u_\mathrm{gut} - \left (\frac{CL}{V_\mathrm{cent}} + \frac{Q}{V_\mathrm{cent}} \right) u_\mathrm{cent} + \frac{Q}{V_\mathrm{peri}} u_\mathrm{peri} \\
  \frac{\mathrm d u_\mathrm{peri}}{\mathrm d t} & = & \frac{Q}{V_\mathrm{cent}} u_\mathrm{cent} - \frac{Q}{V_\mathrm{peri}} u_\mathrm{peri}
\end{eqnarray}
\label{eq:twocpt}
\end{subequations}
with
\begin{itemize}
  \setlength\itemsep{0em}
  \item $u(t)$: drug amount in each compartment (mg),
  \item $k_a$: absorption rate constant (h$^{-1}$),
  \item $CL$: elimination clearance from the central compartment (L / h),
  \item $Q$: intercompartmental clearance (L/h),
  \item $V_\mathrm{cent}$: volume of the central compartment (L),
  \item $V_\mathrm{peri}$: volume of the peripheral compartment (L).
\end{itemize}

{
Both intervention and measurement events are described by the event
schedule. Stan does not have any reserved variable names, but in this
tutorial we follow the NONMEM convention to specify events using the
variable names in Table~\ref{tab:event_schedule}. More details can be found in the \textit{Torsten User Manual}.
}

\begin{table}
  \renewcommand{\arraystretch}{1.5}
  \begin{center}
  \begin{tabular} {l l l}
  \rowcolor[gray]{0.95} \textbf{Variable} & \textbf{Description} \\
  \texttt{cmt} & Compartment in which event occurs.\\
  \rowcolor[gray]{0.95} \texttt{evid} & Type of event: (0) measurement, (1) dosing. \\
  \texttt{addl} & For dosing events, number of additional doses.  \\
  \rowcolor[gray]{0.95} \texttt{ss} & Steady state indicator: (0) no, (1) yes. \\
  \texttt{amt} & Amount of drug administered. \\
  \rowcolor[gray]{0.95} \texttt{time} & Time of the event. \\
  \texttt{rate} & For dosing by infusion, rate of infusion. \\
  \rowcolor[gray]{0.95} \texttt{ii} & For events with multiple dosing, inter-dose interval.
  \end{tabular}
  \end{center}
  \caption{Variables used specify an event schedule.}
  \label{tab:event_schedule}
\end{table}

\subsection{Statistical model}

{
Given a treatment, $x$, and the physiological parameters, $\{k_a, Q,
CL, V_\mathrm{cent}, V_\mathrm{peri} \}$, we compute the drug amounts
$u$ by solving the two--compartment ODE.
We use $y$ to denote measured drug concentration, and $\hat c(=u/V_{\text{cent}})$ the
model--predicted drug concentration. We model the residual error from $y$ to
$\hat{c}$ using a lognormal distribution
}
{
\begin{equation*}
  y \mid \hat c, \sigma \sim \mathrm{logNormal}(\log \hat c, \sigma),
\end{equation*}}
where $\sigma$ is a scale parameter we wish to estimate.
The deterministic computation of $\hat c$ along with the measurement
model, defines our likelihood function $p(c \mid \theta, x)$, where
$\theta = \{k_a, CL, Q, V_\mathrm{cent}, V_\mathrm{peri}, \sigma \}$
and $x$ are input data, i.e. the clinical event schedule. Note that
we are not limited to the above simple model. Stan is capable of
many distributions \cite{Stan_func_ref:2021} as well as encoding more complex residual models
such as the proportional and additive error variance.

It remains to define a prior distribution, $p(\theta)$.
Our prior should allocate probability mass to every plausible parameter value and exclude patently absurd values.
For example the volume of the central compartment is on the order of ten liters, but it cannot be the size of the Sun.
In this simulated example, our priors for the individual parameters are based on population estimates from previous (hypothetical) studies.
\begin{eqnarray*}
  CL & \sim & \mathrm{logNormal}(\log(10), 0.25);  \\
  Q & \sim & \mathrm{logNormal}(\log(15), 0.5); \\
  V_\mathrm{cent} & \sim & \mathrm{logNormal}(\log(35), 0.25); \\
  V_\mathrm{peri} & \sim & \mathrm{logNormal}(\log(105), 0.5); \\
  k_a & \sim & \mathrm{logNormal}(\log(2.5), 1); \\
  \sigma & \sim & \mathrm{Half-Normal}(0, 1); \\
\end{eqnarray*}

Suggestions for building priors can be found in references \cite{Gabry:2019, Betancourt:2020} and at \url{https://github.com/stan-dev/stan/wiki/Prior-Choice-Recommendations}.

\subsection{Specifying a model in Stan}

We can now specify our statistical model using a Stan file, which is divided into coding blocks, each with a specific role.
From R, we then run inference algorithms which take this Stan file as an input.

\subsubsection{Data and parameters block}

To define a model, we need a procedure which returns the log joint distribution, $\log p(\mathcal D, \theta)$.
Our first task is to declare the data, $\mathcal D$, and the parameters, $\theta$, using the coding blocks \texttt{data} and \texttt{parameters}.
It is important to distinguish the two.
The data is fixed.
By contrast, the parameter values change as HMC explores the parameter space, and gradients of the joint density are computed with respect to $\theta$, but not $\mathcal D$.

For each variable we introduce, we must declare a type and, for containers such as arrays, vectors, and matrices, the size of the container\cite[Ch.5]{Stan_users_guide:2021}.
In addition, each statement ends with a semi-colon.
It is possible to specify constraints on the parameters, using the keywords \texttt{lower} and \texttt{upper}.
If one of these constraints is violated, Stan returns an error message.
More importantly, constrained parameters are transformed into unconstrained parameters -- for instance, positive variables are put on the log scale -- which greatly improves computation.

\begin{lstlisting}[style=stan, numbers=none] 
  data {
    int<lower = 1> nEvent;      // number of events
    int<lower = 1> nObs;        // number of observations
    int<lower = 1> iObs[nObs];  // index of events which
                                // are observations.

    // Event schedule
    int<lower = 1> cmt[nEvent];
    int evid[nEvent];
    int addl[nEvent];
    int ss[nEvent];
    real amt[nEvent];
    real time[nEvent];
    real rate[nEvent];
    real ii[nEvent];

    // observed drug concentration
    vector<lower = 0>[nObs] cObs;
  }
  
  parameters {
    real<lower = 0> CL;
    real<lower = 0> Q;
    real<lower = 0> VC;
    real<lower = 0> VP;
    real<lower = 0> ka;
    real<lower = 0> sigma;
  }
\end{lstlisting}

\subsubsection{model block}

Next, the \texttt{model} block allows us to modify the variable \texttt{target}, which Stan recognizes as the log joint distribution.
The following statement increments \texttt{target} using the prior on $\sigma$, which is a normal density, truncated at 0 to only put mass on positive values.
\begin{lstlisting}
target += normal_lpdf(sigma | 0, 1);
\end{lstlisting}
The truncation is implied by the fact $\sigma$ is declared as lower-bounded by 0 in the parameters block.
An alternative syntax is the following:
\begin{lstlisting}
sigma ~ normal(0, 1);
\end{lstlisting}
This statement now looks like our statistical formulation and makes the code more readable.
But we should be mindful that this is not a sampling statement, rather instructions on how to increment \texttt{target}.
We now give the full model block:
\begin{lstlisting}
model {
  // priors
  CL ~ lognormal(log(10), 0.25); 
  Q ~ lognormal(log(15), 0.5);
  VC ~ lognormal(log(35), 0.25);
  VP ~ lognormal(log(105), 0.5);
  ka ~ lognormal(log(2.5), 1);
  sigma ~ normal(0, 1);

  // likelihood
  cObs ~ lognormal(log(concentrationHat[iObs]), sigma);
}
\end{lstlisting}
The likelihood statement involves a crucial term we have not defined yet: \texttt{concentrationHat}.
Additional variables can be created using the \texttt{transformed data} and \texttt{transformed parameters} blocks.
We will take advantage of these to compute the drug concentration in the central compartment for each event.
Note that for the likelihood, we only use the concentration during observation events, hence the indexing \texttt{[iObs]}.

\subsubsection{Transformed data and transformed parameters block} \label{sec:twocpt_transformed_parameters}

In \texttt{transformed data}, we can construct variables which only depend on the data.
For this model, we simply specify the number of compartments in our model (including the gut), \texttt{nCmt}, and the numbers of pharmacokinetic parameters, \texttt{nTheta}, two variables which will come in handy shortly.
\begin{lstlisting}
  transformed data {
    int nCmt = 3;
    int nTheta = 5; 
  }
\end{lstlisting}
Because the data is fixed, this operation is only computed once.
By contrast, operations in the \texttt{transformed parameters} block need to be performed (and differentiated) for each new parameter values.

To compute \texttt{concentrationHat} we need to solve the relevant ODE within the clinical event schedule.
Torsten provides a function which returns the drug mass in each compartment at each time point of the event schedule.
\begin{lstlisting}
matrix<lower = 0>[nCmt, nEvent] 
  mass = pmx_solve_twocpt(time, amt, rate, ii, evid,
                          cmt, addl, ss, theta);
\end{lstlisting}
The first eight arguments define the event schedule and the last argument, \texttt{theta}, is an array containing the pharmacokinetic parameters, and defined as follows:
\begin{lstlisting}
real theta[nTheta] = {CL, Q, VC, VP, ka};
\end{lstlisting}
It is also possible to have \texttt{theta} change between events, and specify lag times and bioavailability fractions, although we will not take advantage of these features in the example at hand.

The Torsten function we have chosen to use solves the ODEs analytically.
Other routines use a matrix exponential, a numerical solver, or a combination of analytical and numerical methods \cite{Margossian:2017}.
It now remains to compute the concentration in the central compartment at the relevant times.
The full \texttt{transformed parameters} block is as follows:
\begin{lstlisting}
transformed parameters {
  real theta[nTheta] = {CL, Q, VC, VP, ka};
  row_vector<lower = 0>[nEvent] concentrationHat;
  matrix<lower = 0>[nCmt, nEvent] mass;

  mass = pmx_solve_twocpt(time, amt, rate, ii, evid, 
                          cmt, addl, ss, theta);

  // Extract mass in central compartment and divide 
  // by central volume.
  concentrationHat = mass[2, ] ./ VC;
}  
\end{lstlisting} 

The Stan file contains all the coding blocks in the following order: \texttt{data}, \texttt{transformed data}, \texttt{parameters}, \texttt{transformed parameters}, \texttt{model}.
The full Stan code can be found in the Supplementary Material.

\subsection{Calling Stan from R}\label{sec:call_stan_from_R}
The R package \texttt{cmdstanr} allows us to run a number of algorithms on a model defined in a Stan file.
An excellent place to get started with the package is \url{https://mc-stan.org/cmdstanr/articles/cmdstanr.html}.

The first step is to ``transpile'' the file -- call it \texttt{twocpt.stan} --, that is translate the file into C++ and then compile it.
\begin{lstlisting}
mod <- cmdstan_model("twocpt.stan")
\end{lstlisting}
We can then run Stan's HMC sampler by passing in the requisite data and providing other tuning parameters.
Here: (i) the number of Markov chains (which we run in parallel), (ii) the initial value for each chain, (iii) the number of warmup iterations, and (iv) the number of sampling iterations.
\begin{lstlisting}
fit <- mod$sample(data = data, chains = n_chains,
                  parallel_chains = n_chains,
                  init = init,
                  iter_warmup = 500, 
                  iter_sampling = 500)
\end{lstlisting}

{
By default, Stan uses 1,000 warmup iterations and 1,000 sampling iterations.
Empirically these defaults work well across a broad range of models when running an adaptive HMC sampler. 
For relatively simple models, we may even use shorter warmup and sampling phases, as we have done above.
This should be contrasted with random walk MCMC, such as the Gibbs sampler in BUGS, where it is typical to run 5,000 or even 10,000 iterations per phase.
Random walk MCMC tends to generate Markov chains with a higher autocorrelation than HMC,
hence the need to run more iterations.
In the next two sections, we will discuss diagnostics which can be used to adjust the length of the warmup and sampling phases.
}

There are several other arguments we can pass to the sampler and which we will take advantage of throughout the tutorial.
For applications in pharmacometrics, we recommend specifying the initial starting points via the \texttt{init} argument, as the defaults may not be appropriate.
In this tutorial, we draw the initial points from their priors by defining an appropriate R function.

The resulting \texttt{fit} object stores the samples generated by HMC from which can deduce the sample mean, sample variance, and sample quantiles of our posterior distribution.
This information is readily accessible using \texttt{fit\$summary()} and summarized in table~\ref{tab:summary}.
We could also extract the samples and perform any number of operations on them.

\begin{table}[!h]
  \renewcommand{\arraystretch}{1.5}
  \begin{tabular}{l l l l l l l l l l}
  \rowcolor[gray]{0.95} & \bf mean & \bf median & \bf sd & \bf mad & \bf q5 & \bf q95 & $\bf \hat R$ & \bf ESS$_\mathrm{bulk}$ & \bf ESS$_\mathrm{tail}$ \\
$CL$    &    10.0  &  10.0  &  0.378  & 0.367 &  9.39 &  10.6  &  1.00   & 1580  &  1348 \\
\rowcolor[gray]{0.95} $Q$      &  19.8    & 19.5  &  4.00  &  4.01 &  13.8  &   26.8 &   1.00  &   985 &    1235 \\
$V_\mathrm{cent}$   &   41.2  &  40.8 &   9.71   & 9.96  & 25.6 &   57.7   & 1.00   &  732  &  1120 \\
\rowcolor[gray]{0.95} $V_\mathrm{peri}$    &  124 &   123 &    18.0  &  18.0 &    97.1 &  155 &   1.00  &  1877 &    1279 \\
$k_a$       &  1.73   & 1.67 &  0.523  & 0.522   & 1.01 &   2.68 &   1.00   &  762 &    1108 \\
\rowcolor[gray]{0.95} $\sigma$   & 0.224 &   0.222 &  0.0244 &  0.0232 &  0.187 &   0.269 &  1.01  & 1549 &   1083
 \end{tabular}
  \caption{Summary of results when fitting a two compartment model. \textit{The first columns return sample estimates of the posterior mean, median, standard deviation, median absolute deviation, $5^\mathrm{th}$ and $95^\mathrm{th}$ quantiles, based on our approximate samples.
  The next three columns return the $\hat R$ statistics and the effective sample size for bulk and tail estimates, and can be used to identify problems with our inference.}}
  \label{tab:summary}
\end{table}

\subsection{Checking our inference}

Unfortunately there is no guarantee that a particular algorithm will work across all the applications we will encounter.
We can however make sure that certain necessary conditions are met.

Much of the MCMC literature focuses on estimating expectation values for quantities of interest $f$,
\begin{equation*}
 \mathbb E f = \int_\Theta f(\theta) p(\theta \mid y) \text d \theta,
\end{equation*}
using sample estimators
\begin{equation*}
  \widehat{\mathbb E} f = \frac{1}{n} \sum_{i = 1}^n f \left (\theta^{(i)} \right),
\end{equation*}
for some samples $\theta^{(1)}, \theta^{(2)}, \cdots, \theta^{(n)}$.
When constructing such estimators using MCMC samples, rather than with exact independent samples, we must account for the fact that our samples are correlated and biased.

\subsubsection{Checking for convergence with $\hat R$} \label{sec:Rhat}

{ MCMC samples are biased because Markov chains generate correlated samples, meaning any sample has some correlation with the initial point.}
If we run the algorithm for enough iterations, the correlation to the initial point becomes negligible and the chain ``forgets'' its starting point.
But what constitutes enough iterations?

\begin{figure}
   \begin{center}
   \includegraphics[width = 6in]{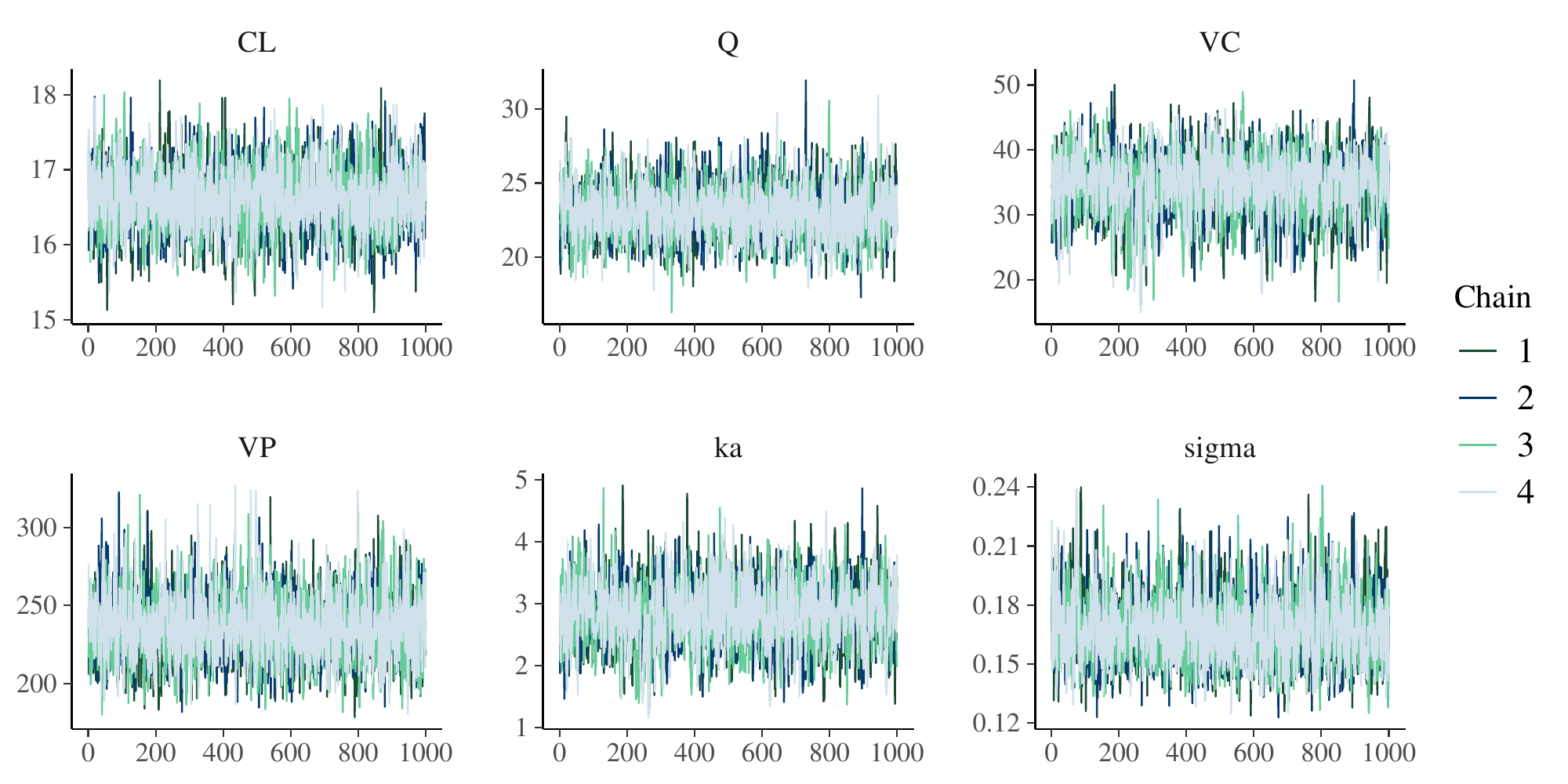}
   \caption{Trace plots. \textit{The sampled values for each parameters are plotted against the iterations during the sampling phase. Multiple Markov chains were initialized at different points. However, once in the sampling phase, we cannot distinguish the chains.}}
   \label{fig:trace}
   \end{center}
 \end{figure}

To monitor bias, we run multiple Markov chains, each started at different points, and check that they all converge to the same region of the parameter space.
One way to check this is to compute the $\hat R$ statistics, for which we provide an intuitive definition:
\begin{equation*}
  \hat R \overset{\mathrm{intuitively}}{=} \frac{\mathrm{Total \ variance \ across \ all \ chains}}{\mathrm{Average \ within \ chain \ variance}}.
\end{equation*}
If the chains are mixing properly, { both the numerator and denominator measure the posterior variance,} and $\hat R$ converges to 1.0, as $n$ increases.
Moreover, we want $\hat R \approx 1.0$, as is the case in table~\ref{tab:summary}.
Stan uses an improved $\hat R$ statistics described in a recent paper by Vehtari et al (2020) \cite{Vehtari:2020}.
We can also visually check that the chains are properly mixing using a trace plot (Figure~\ref{fig:trace}).


If $\hat R \gg 1$ and, more generally, if the chains were not mixing, this would be cause for concern and an invitation to adjust our inference method.
{ One potential solution is to increase the warmup length.}
Even when $\hat R \approx 1$, we should entertain the possibility that all the chains suffer from the same bias.

\subsubsection{Controlling the variance of our estimator}\label{sec:variance_control}

Let's assume that our warmup phase is long enough and the bias negligable.
The expected error of our sample estimator is now determined by the variance.
Under certain regularity conditions, our estimator follows an \textit{MCMC central limit theorem},
{
\begin{equation} \label{eq:CLT}
  \hat {\mathbb E} f \overset{\mathrm{approx}}{\sim} \mathrm{Normal} \left ( \mathbb E f, \frac{\sigma_f}{\sqrt{n_\mathrm{eff}}} \right )
\end{equation}}
where $n_\mathrm{eff}$ is the \textit{effective sample size}, denoted $\text{ESS}_\text{bulk}$ in Table~\ref{tab:summary}.
Deviations from this approximation have order $\mathcal O \left (1 / n_\mathrm{eff}^2 \right)$.
In the limiting case where we generate independent samples, $n_\mathrm{eff} = n$;
however, when samples exhibits correlation, $n_\mathrm{eff} < n$
and the variance of our sample estimator increases.
%
For $CL$, we have 2,000 samples, but the effective sample size is 1,580 (Table~\ref{tab:summary}).
If $n_\mathrm{eff}$ is low, our estimator may not be precise enough
{
and we should increase the sampling phase to generate more samples.
}

{
Achieving $n_\text{eff} \approx 100$ is, in our experience, usually sufficient in an applied setting.
This means that the variance of the sample estimator is 1\% that of the posterior, as can be seen from Equation~\eqref{eq:CLT}.
At this point the uncertainty is dominated by the intrinsic posterior variance rather than the error in our inference procedure.
If the effective sample size is below 100 for certain quantities, Stan issues a warning message.
}


The effective sample size is only formally defined in the context of estimators for expectation values.
We may also be interested in tail quantities, such as extreme quantiles, which are more difficult to estimate and require many more samples to achieve a desired precision.
Vehtari et al (2020) \cite{Vehtari:2020} propose a generalization of the effective sample size for such quantities, and introduce the \textit{tail effective sample size}.
This is to be distinguished from the traditional effective sample size, henceforth the \textit{bulk effective sample size}.
Both quantities are reported by Stan.

\subsection{Checking the model: posterior predictive checks}  \label{sec:twoCpt_ppc}

Once we develop enough confidence in our inference, we still want to check our fitted model.
There are many ways of doing this.
We may look at the posterior distribution of an interpretable parameter and see if it suggests implausible values.
Or we may evaluate the model's ability to perform a certain task, e.g. classification or prediction, as is often done in machine learning.
In practice, we find it useful to do \textit{posterior predictive
  checks} (PPC), that is simulate data from the fitted model and compare the simulation to the observed data \cite[chapter 6]{Gelman:2013}.
Mechanically, the procedure is straightforward:
\begin{enumerate}
  \item Draw the parameters from their posterior, $\tilde \theta \sim p(\theta \mid y).$
  \item Draw predicted observations from the likelihood, conditional on the drawn parameters, $\tilde y \sim p(y \mid \tilde \theta)$.
\end{enumerate}
This amounts to drawing observations from their posterior distribution, that is $\tilde y \sim p(\tilde y \mid y)$.
Both the uncertainty due to our estimation and the uncertainty due to our measurement model propagate to our predictions.

\begin{figure}
   \begin{center}
   \includegraphics[width = 6in]{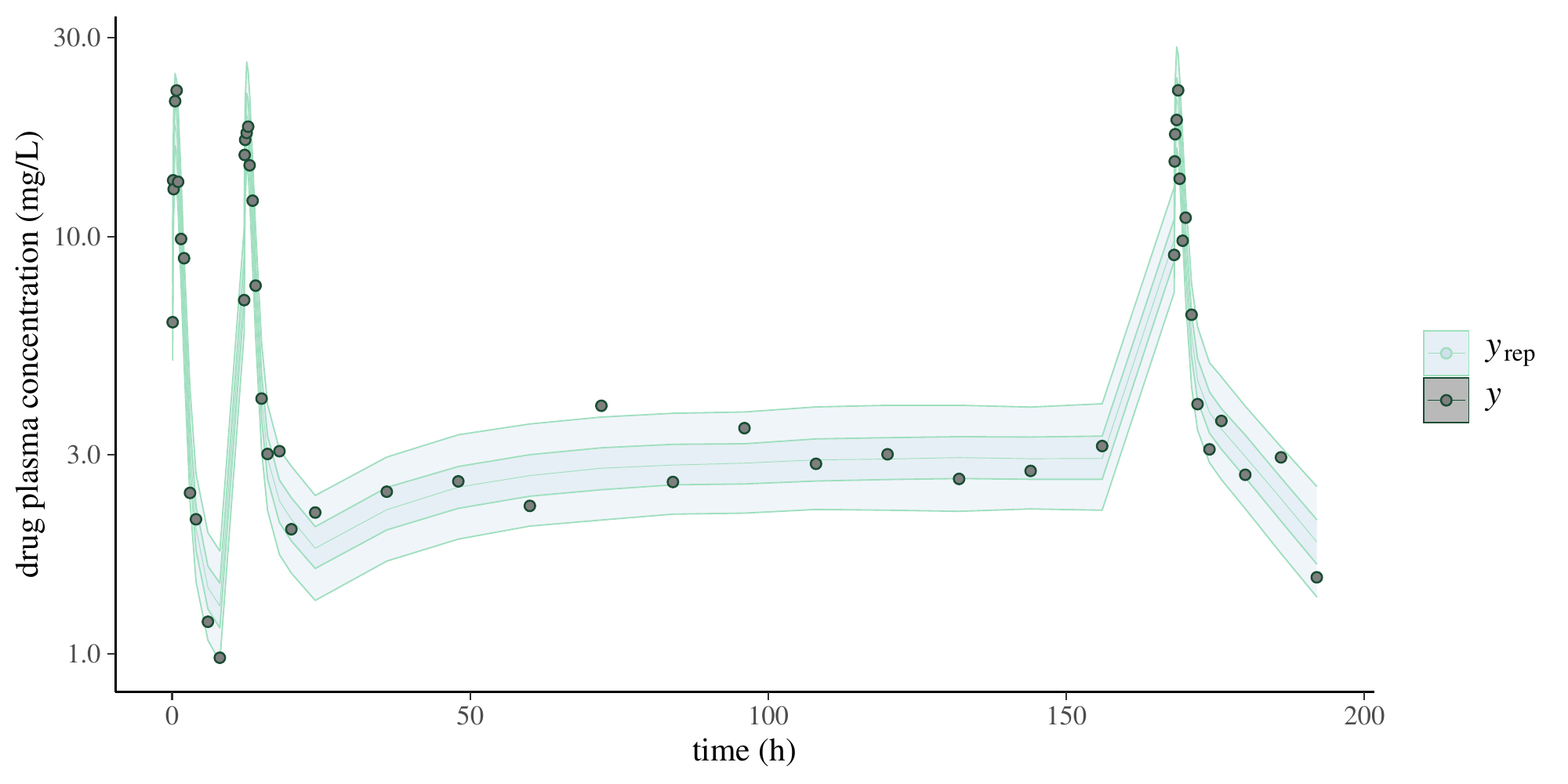}
   \end{center}
   \caption{Posterior predictive checks for two compartment model. \textit{The circles represent the observed data and the shaded areas the $50^\mathrm{th}$ and $90^\mathrm{th}$ credible intervals based on posterior draws.}}
   \label{fig:ppc} 
 \end{figure}

Stan provides a \texttt{generated quantities} block, which allows us to compute values, based on sampled parameters.
In our two compartment model example, the following code draws predicted observations from the likelihood:
\begin{lstlisting}
generated quantities {
  real concentrationObsPred[nObs] 
    = lognormal_rng(log(concentrationHat[iObs]), sigma);
}
\end{lstlisting}
We generate predictions at the observed points for each sampled point, $\theta^{(i)}$.
This gives us a sample of predictions and we can use the $5^\mathrm{th}$ and $95^\mathrm{th}$ quantiles to construct a credible interval.
We may then plot the observations and the credible intervals (Figure~\ref{fig:ppc}) and see that, indeed, the data generated by the model is consistent with the observations.


\subsection{Comparing models: leave-one-out cross validation}

\begin{figure}
   \begin{center}
   \includegraphics[width = 6in]{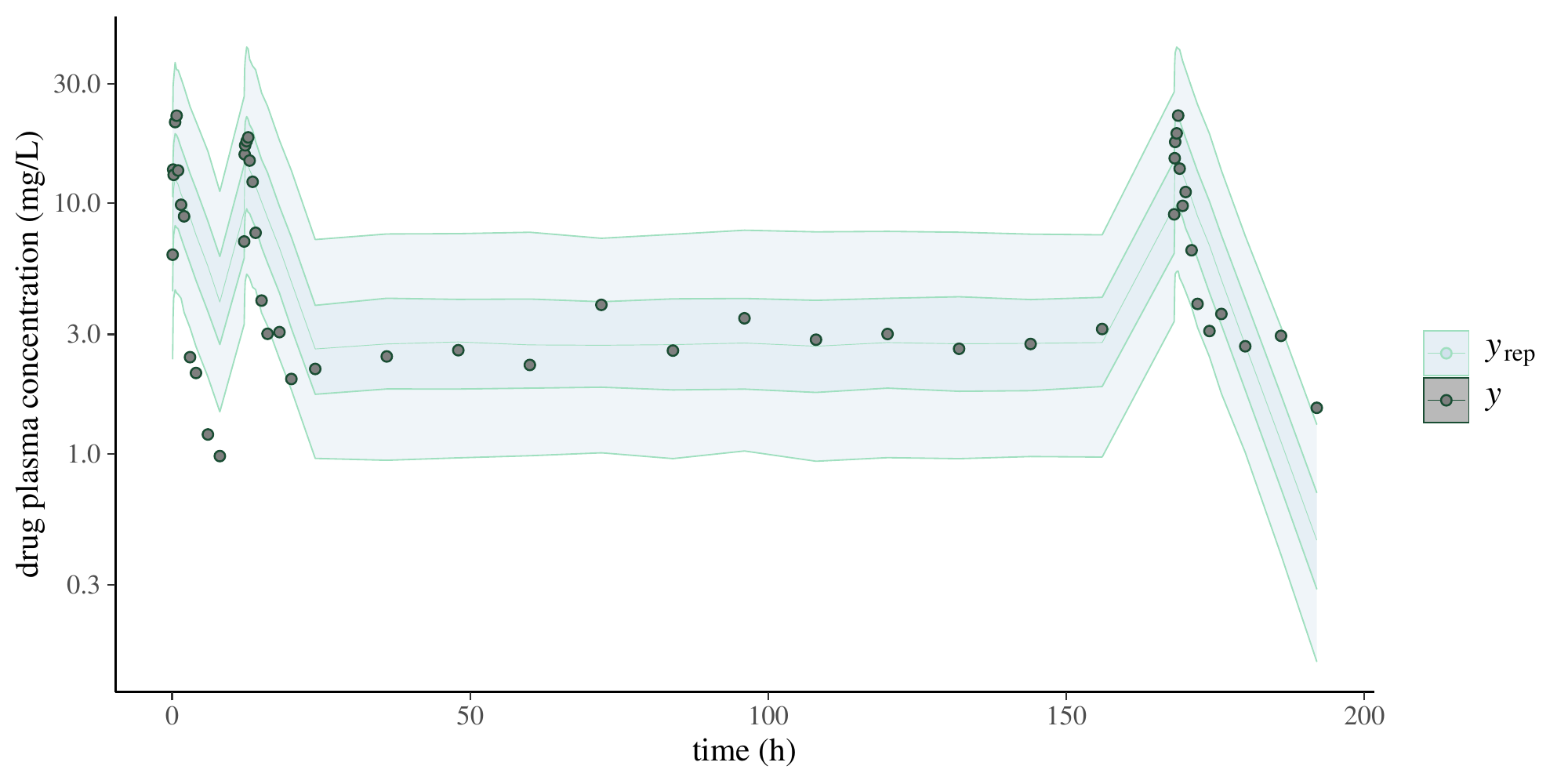}
   \end{center}
   \caption{Posterior predictive checks for one compartment model. \textit{The circles represent the observed data and the shaded areas the $50^\mathrm{th}$ and $90^\mathrm{th}$ credible intervals based on posterior draws. A graphical inspection suggests the credible interval is wider for the one compartment model than they are for the two compartment model.}}
   \label{fig:ppc_onecpt} 
 \end{figure}

Beyond model criticism, we may be interested in model comparison.
Continuing our running example, we compare our two compartment model to a one compartment model, which is also supported by Torsten via the \texttt{pmx\_solve\_onecpt} routine.
The corresponding posterior predictive checks are shown in Figure~\ref{fig:ppc_onecpt}.


There are several ways of comparing models and which method is appropriate crucially depends on the insights we wish to gain.
If our goal is to asses a model's ability to make good out-of-sample predictions, we may consider \textit{Bayesian leave-one-out} (LOO) cross validation.
The premise of cross-validation is to exclude a point, $(y_i, x_i)$, from the \textit{training set}, i.e. the set of data to which we fit the model.
Here $x_i$ denotes the covariate and in our example, the relevant row in the event schedule.
We denote the reduced data set, $y_{-i}$.
We then generate a prediction $(\tilde y_i, x_i)$ using the fitted model, and compare $\tilde y_i$ to $y_i$.
A classic metric to make this comparison is the squared error, $(\tilde y_i - y_i)^2$.

Another approach is to use the \textit{LOO estimate of out-of-sample predictive fit}:
\begin{equation*}
  \mathrm{elp}_\mathrm{loo} := \sum_{i}^n \log p(y_i \mid y_{-i}).
\end{equation*}
Here, no prediction is made.
We however examine how consistent an ``unobserved'' data point is with our fitted model.
Computing this estimator is expensive, since it requires fitting the model to $n$ different training sets in order to evaluate each term in the sum.

Vehtari et al (2016) \cite{Vehtari:2016} propose an estimator of $\mathrm{elp}_\mathrm{loo}$, which uses Pareto smooth importance sampling and only requires a single model fit.
The premise is to compute
\begin{equation*}
  \log p(y_i \mid y)
\end{equation*}
and correct this value, using importance sampling, to estimate $\log p(y_i \mid y_{-i})$.
Naturally this estimator may be inaccurate.
What makes this tool so useful is that we can use the Pareto shape parameter, $\hat k$, to asses how reliable the estimate is.
In particular, if $\hat k > 0.7$, then the estimate shouldn't be trusted.
The estimator is implemented in the R package \texttt{loo}. see
\cite{Gabry:2020} for more details including its connection and
comparison to the widely applicable information criterion (WAIC).

\begin{figure}
   \begin{center}
     \includegraphics[width = 5in]{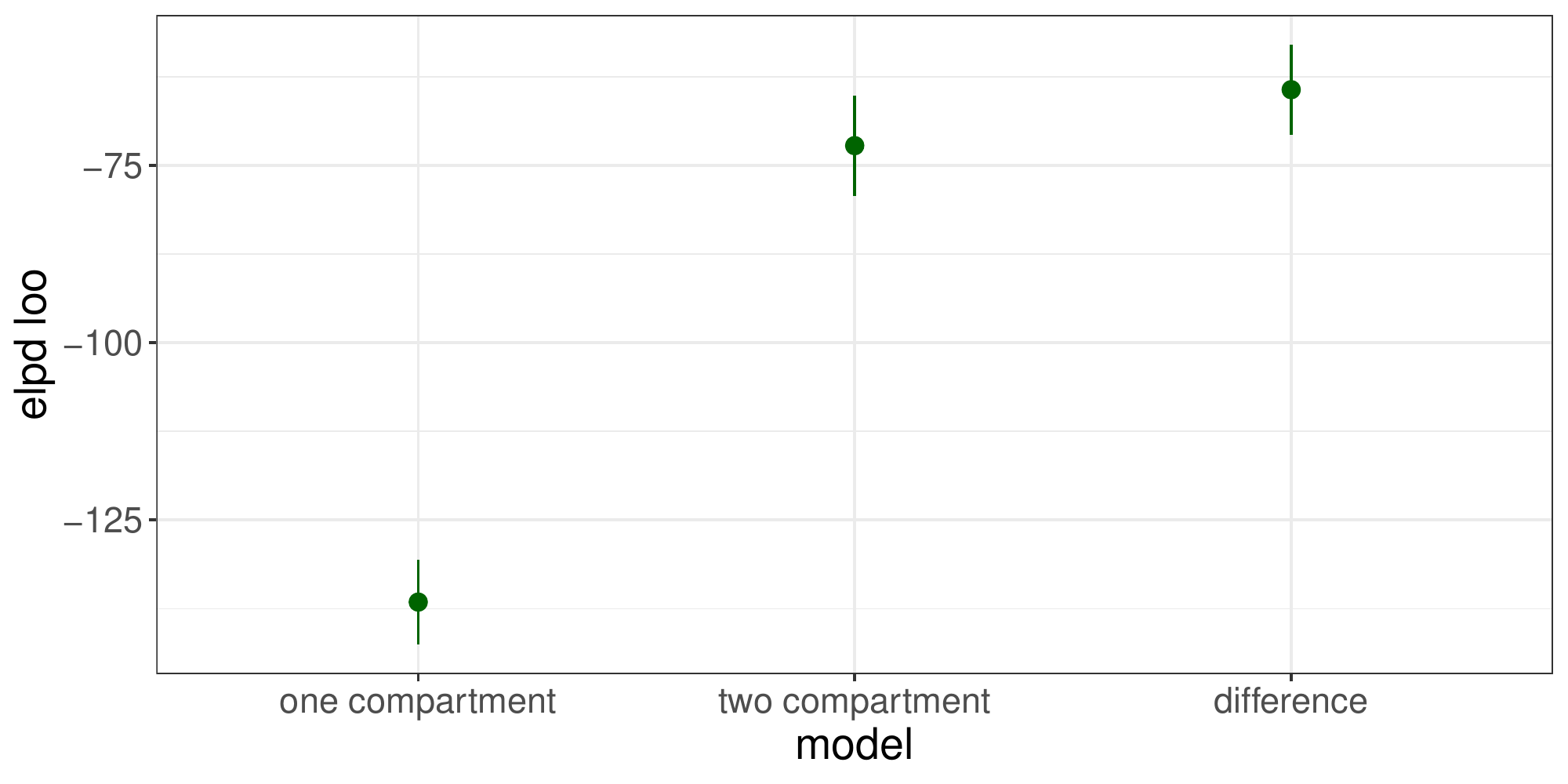}
     \caption{Leave-one-out estimate of out-of-sample predictive
       fit. \textit{Plotted is the estimate,
         $\mathrm{elp}_\mathrm{loo}$, for the one and two compartment
         models as well as the difference in elpd for the two models. Clearly, the two compartment model has superior predictive capabilities.}}
     \label{fig:loo}
   \end{center}
 \end{figure}

Conveniently, we can compute $\log p(y_i \mid y)$ in Stan's \texttt{generated quantities} block.
\begin{lstlisting}
vector[nObs] log_lik;
for (i in 1:nObs)
  log_lik[i] = 
   lognormal_lpdf(cObs[i] | log(concentrationHat[iObs[i]]),
                  sigma);
\end{lstlisting}
These results can then be extracted and fed into Loo to compute $\mathrm{elp}_\mathrm{loo}$.
The file \texttt{twoCpt.r} in the Supplementary Material shows exactly how to do this.
Figure~\ref{fig:loo} plots the estimated $\mathrm{elp}_\mathrm{loo}$, along with a standard deviation, and shows the two compartment model has better out-of-sample predictive capabilities.


\section{Two compartment population model}

We now consider the scenario where we have data from multiple patients and fit a population model.
Population models are a powerful tool to capture the heterogeneity between patients, while also recognizing similarities.
Building the right prior allows us to pool information between patients, the idea being that what we learn from one patient teaches us something -- though not everything -- about the other patients.
In practice, such models can frustrate inference algorithms and need
to be implemented with care \cite{Betancourt:2013}.
We start with an example where the interaction between the model and our MCMC sampler is well behaved.
In Part II of this tutorial, we will examine a more difficult case, for which we will leverage Stan's diagnostic capabilities in order to run reliable inference.

\subsection{Statistical model} \label{sec:twoCptPop_stat}

Let $\vartheta$ be the 2D array of body weight-normalized pharmacokinetic parameters for each patient,
with
\begin{equation*}
  \vartheta_j = (CL_{\mathrm{norm}, j}, Q_{\mathrm{norm}, j}, V_{\mathrm{cent}, \mathrm{norm},  j}, V_{\mathrm{peri}, \mathrm{norm}, j}, k_{a, j}),
\end{equation*} 
the parameters for the $j^\mathrm{th}$ patient.
We construct a population model by introducing random
variation to describe otherwise unexplained inter-individual
variability. In a Bayesian context this is sometimes referred to as a
prior distribution for the individual parameters,
\begin{eqnarray*}
  \vartheta_j \sim \mathrm{LogNormal} (\log \vartheta_\mathrm{pop}, \Omega).
\end{eqnarray*}
As before we work on the log scale to account for the fact the pharmacokinetic parameters are constrained to be positive.
$\vartheta_\mathrm{pop} = (CL_\mathrm{pop}, Q_\mathrm{pop}, V_\mathrm{cent, pop}, V_\mathrm{peri, pop}, k_\mathrm{a, pop})$ is the population mean (on the logarithmic scale) and $\Omega$ the population covariance matrix.
Both $\vartheta_\mathrm{pop}$ and $\Omega$ are estimated.
In this example, we start with the simple case where $\Omega$ is
diagonal.
For our example we will also use conventional allometric scaling to
adjust the clearance and volume parameters for body weight.

{
  \begin{align*}
    CL_j &= CL_{\mathrm{norm}, j} \left( \frac{\mathrm{weight}}{70} \right)^{0.75} = \vartheta_{1j}\left(  \frac{\mathrm{weight}}{70}\right)^{0.75} \\
    Q_j &= Q_{\mathrm{norm}, j} \left( \frac{\mathrm{weight}}{70}\right)^{0.75}  = \vartheta_{2j} \left(\frac{\mathrm{weight}}{70}\right)^{0.75} \\
    V_{\mathrm{cent}, j} &= V_{\mathrm{cent}, \mathrm{norm},  j} \frac{\mathrm{weight}}{70} = \vartheta_{3j} \frac{\mathrm{weight}}{70} \\
    V_{\mathrm{peri}, j} &= V_{\mathrm{peri}, \mathrm{norm},  j} \frac{\mathrm{weight}}{70} = \vartheta_{4j} \frac{\mathrm{weight}}{70}
  \end{align*}
}

The likelihood remains mostly unchanged, with the caveat that it must now be computed for each patient.
Putting this all together, we have the following model, as specified by the joint distribution,
\begin{eqnarray*}
  \vartheta_\mathrm{pop} & \sim & p(\vartheta_\mathrm{pop}), \hspace{1in} \text{(prior on pharmacokinetic parameters)} \\
  \Omega & \sim & p(\Omega), \hspace{1.2in} \text{(prior on population covariance)} \\
  \sigma & \sim & p(\sigma) \\
  \vartheta \mid \vartheta_\mathrm{pop}, \Omega  & \sim  & \mathrm{logNormal}(\log \vartheta_\mathrm{pop}, \Omega), \\
  y \mid c, \sigma & \sim & \mathrm{LogNormal}(\log c, \sigma).
\end{eqnarray*}

\subsection{Specifying the model in Stan}

We begin by adjusting our parameters block:
\begin{lstlisting}[style=stan, basicstyle=\ttfamily\footnotesize, numbers=none]
parameters {
  // Population parameters
  real<lower = 0> CL_pop;
  real<lower = 0> Q_pop;
  real<lower = 0> VC_pop;
  real<lower = 0> VP_pop;
// Constrain ka_pop > lambda_1 
  real<lower = (CL_pop / VC_pop + Q_pop / VC_pop + Q_pop / VP_pop +
                sqrt((CL_pop / VC_pop + Q_pop / VC_pop +
                  Q_pop / VP_pop)^2 -
                4 * CL_pop / VC_pop * Q_pop / VP_pop)) / 2 >
                   ka_pop;
 
  // Inter-individual variability
  vector<lower = 0>[nIIV] omega;
  real<lower = 0> theta[nSubjects, nTheta];

  // Residual variability
  real<lower = 0> sigma;
}
\end{lstlisting}
The declaration for $k_\mathrm{a, pop}$ illustrates that constraints may be
expressions including other variables in the model. In this case
$k_\mathrm{a, pop}$ is constrained to avoid identifiability problems
due to ''flip-flop''.

The variable, $\vartheta_\mathrm{pop}$ is introduced in \texttt{transformed parameters},
mostly for convenience purposes:
\begin{lstlisting}[style=stan, numbers=none]
vector<lower = 0>[nTheta]  
  theta_pop = to_vector({CL_pop, Q_pop, VC_pop, VP_pop, 
                         ka_pop});
\end{lstlisting}
The model block reflects our statistical formulation:
\begin{lstlisting}[style=stan, numbers=none] 
model {
  // prior on population parameters
  CL_pop ~ lognormal(log(10), 0.25); 
  Q_pop ~ lognormal(log(15), 0.5);
  VC_pop ~ lognormal(log(35), 0.25);
  VP_pop ~ lognormal(log(105), 0.5);
  ka_pop ~ lognormal(log(2.5), 1);

  // independent lognormal priors on each element of omega.
  omega ~ lognormal(0.25, 0.1);
  
  sigma ~ normal(0, 1);

  // interindividual variability
  for (j in 1:nSubjects)
    theta[j, ] ~ lognormal(log(theta_pop), omega);

  // likelihood
  cObs ~ lognormal(log(concentrationObs), sigma);
}
\end{lstlisting}

In the \texttt{transformed parameters} block we also declare and
calculate the individual parameters given $\vartheta_j$ and any
relevant covariates---body weight in this case.
\begin{lstlisting}[style=stan, numbers=none] 
  // Individual parameters
  vector<lower = 0>[nSubjects] CL = to_vector(theta[, 1]) .* exp(0.75 * log(weight / 70 ));
  vector<lower = 0>[nSubjects] Q = to_vector(theta[, 2]) .* exp(0.75 * log(weight / 70 ));
  vector<lower = 0>[nSubjects] VC = to_vector(theta[, 3]) .* (weight / 70);
  vector<lower = 0>[nSubjects] VP = to_vector(theta[, 4]) .* (weight / 70);
  vector<lower = 0>[nSubjects] ka = to_vector(theta[, 5]);
\end{lstlisting}

It remains to compute \texttt{concentrationObs}.
There are several ways to do this and, depending on the computational resources available, we may either compute the concentration for each patients sequentially or in parallel.
For now, we do the simpler sequential approach.
In the upcoming Part II of this tutorial, we examine how Torsten offers easy-to-use parallelization  for population models.

Sequentially computing the concentration is a simple matter of bookkeeping.
In \texttt{transformed parameters} we loop through the patients using a \texttt{for} loop.
The code is identical to what we used in Section~\ref{sec:twocpt_transformed_parameters},
with the caveat that the arguments to \texttt{pmx\_solve\_twocpt} are now indexed to indicate for which patient we compute the drug mass.
For example, assuming the time schedule is ordered by patient, the event times corresponding to the $j^\mathrm{th}$ patient are given by
\begin{lstlisting}[style=stan, numbers=none]
time[start[j]:end[j]]
\end{lstlisting}
where \texttt{start[j]} and \texttt{end[j]} contain the indices of the first and last event for the $j^\mathrm{th}$ patient, and the syntax for indexing is as in R.
The full \texttt{for} loop is then
\begin{lstlisting}[style=stan, numbers=none]
for (j in 1:nSubjects) {
  mass[, start[j]:end[j]] =
       pmx_solve_twocpt(time[start[j]:end[j]],
                        amt[start[j]:end[j]],
                        rate[start[j]:end[j]],
                        ii[start[j]:end[j]],
                        evid[start[j]:end[j]],
                        cmt[start[j]:end[j]],
                        addl[start[j]:end[j]],
                        ss[start[j]:end[j]],
                        {CL[j], Q[j], VC[j], VP[j], ka[j]});

  concentration[start[j]:end[j]] = 
                    mass[2, start[j]:end[j]] / VC[j];
}
\end{lstlisting}
Note that the last vector argument in \texttt{pmx\_solve\_twocpt} is
generated using \texttt{\{\}} syntax.

Once we have written our Stan model, we can apply the same methods for inference and diagnostics as we did in the previous section.

\subsection{Posterior predictive checks}

\begin{figure}
   \begin{center}
   \includegraphics[width=6in]{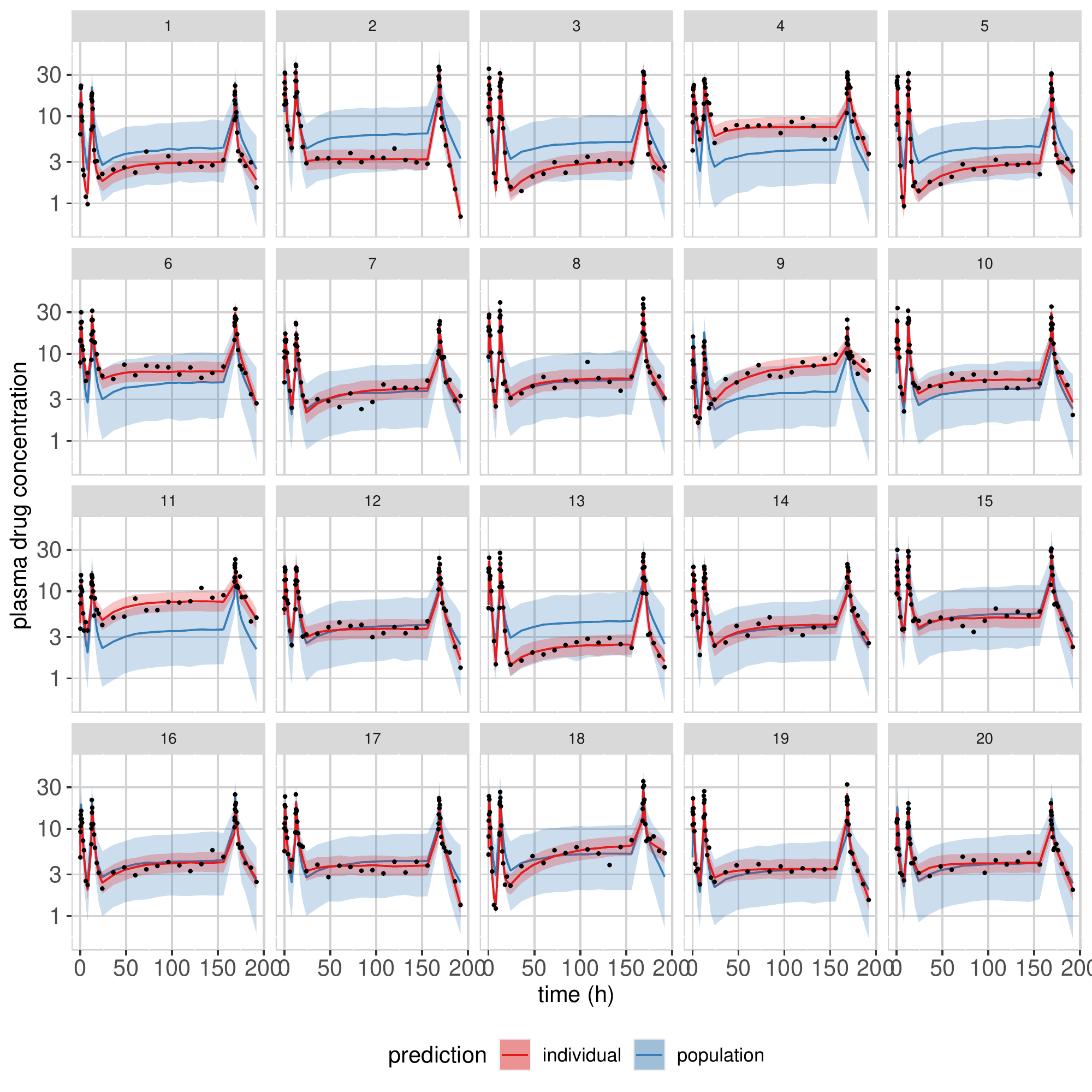}
   \end{center}
   \caption{Population two compartment model: Posterior predictive
     checks for each individual. \textit{Key: Black dots = observed data; red
     curve and shaded area = posterior median and 90\% credible
     intervals for prediction of new observations in the same
     individual; blue curve and shaded area = posterior median and 90\% credible
     intervals for prediction of new observations in a hypothetical new
     individual with the same body weight.}}
   \label{fig:twoCptPop_ppc}
 \end{figure}

\begin{figure}
   \begin{center}
   \includegraphics[width=6in]{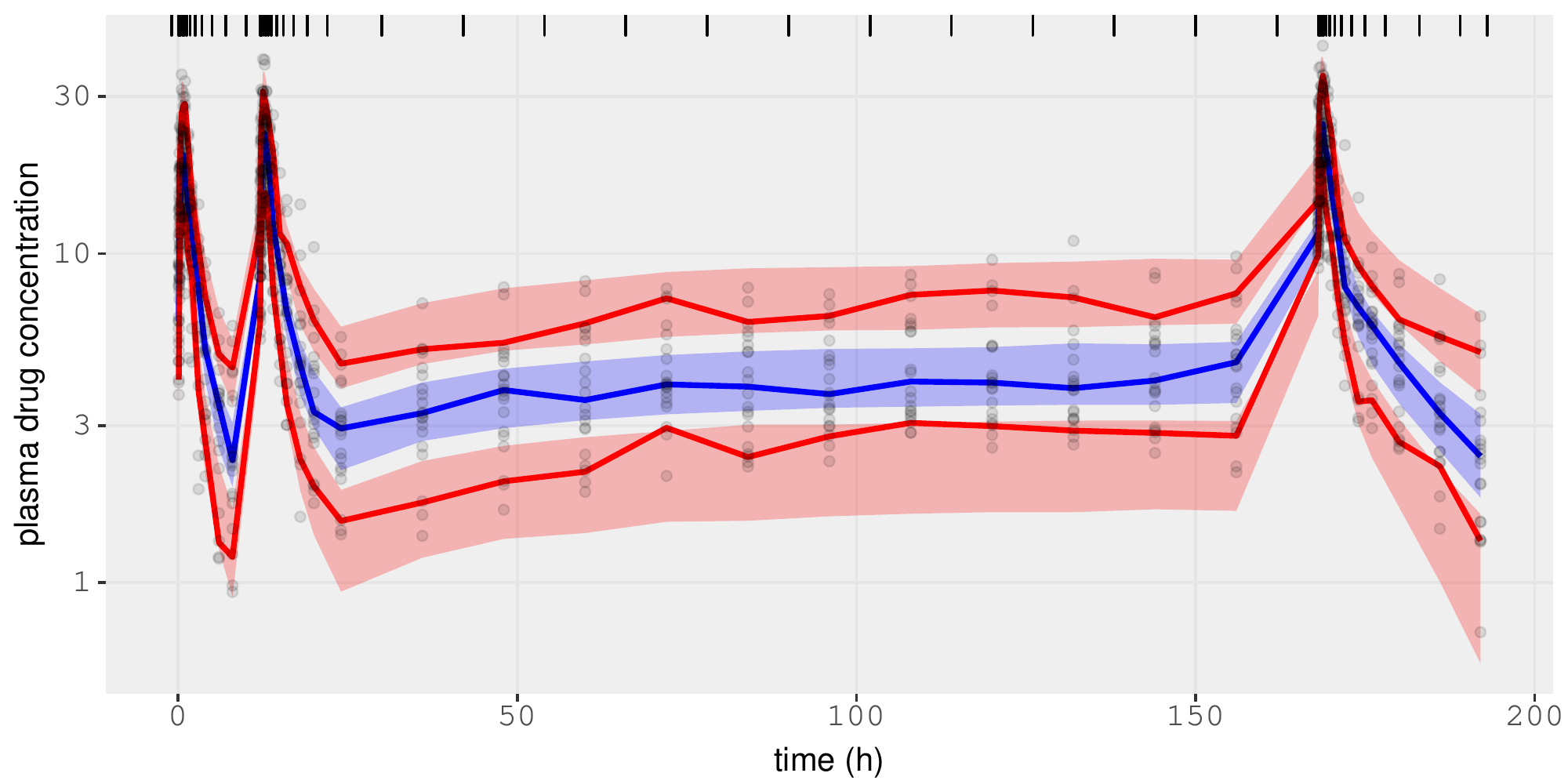} 
   \caption{Population two compartment model: Posterior predictive
     checks for all individuals. \textit{Key: Black circles = observed data; blue
     curves and shaded areas = posterior median and 80\% credible
     intervals for the population median; red curve and shaded area =
     posterior median and 80\% credible intervals for the $10^\text{th}$ and $90^\text{th}$
     population percentiles
     intervals }}
     \end{center}
   \label{fig:twoCptPop_ppc_all}
 \end{figure}

We follow the exact same procedure as in Section~\ref{sec:twoCpt_ppc}
-- using even the same line of code -- to simulate new observations
for the same patients we analyzed.
Figure~\ref{fig:twoCptPop_ppc} plots posterior predictions for each
individual patient.
In addition, we simulate new observations for hypothetical new
patients by: (i) drawing pharmacokinetic parameters from our 
population distribution, (ii) solving the ODEs with these simulated
parameters and (iii) using our measurement model to simulate new
observations.
Those predictions are also shown in Figure~\ref{fig:twoCptPop_ppc} for
each individual. Figure~\ref{fig:twoCptPop_ppc_all} depicts a composite posterior predictive check for all
individuals.
The generated quantities block then looks as follows:
%
\begin{lstlisting}[style=stan, numbers=none]
generated quantities {
  real concentrationObsPred[nObs] 
    = exp(normal_rng(log(concentration[iObs]), sigma));

  real cObsPred[nObs];
  matrix<lower = 0>[nCmt, nEvent] massPred;
  real thetaPred[nSubjects, nTheta];
  row_vector<lower = 0>[nEvent] concentrationPred;
  vector<lower = 0>[nSubjects] CLPred;
  vector<lower = 0>[nSubjects] QPred;
  vector<lower = 0>[nSubjects] VCPred;
  vector<lower = 0>[nSubjects] VPPred;
  vector<lower = 0>[nSubjects] kaPred;

  for (j in 1:nSubjects) {
    thetaPred[j, ] = lognormal_rng(log(theta_pop), omega);

  CLPred = to_vector(thetaPred[, 1]) .* exp(0.75 * log(weight / 70 ));
  QPred = to_vector(thetaPred[, 2]) .* exp(0.75 * log(weight / 70 ));
  VCPred = to_vector(thetaPred[, 3]) .* (weight / 70);
  VPPred = to_vector(thetaPred[, 4]) .* (weight / 70);
  kaPred = to_vector(thetaPred[, 5]);

  massPred[, start[j]:end[j]]
      = pmx_solve_twocpt(time[start[j]:end[j]],
                      amt[start[j]:end[j]],
                      rate[start[j]:end[j]],
                      ii[start[j]:end[j]],
                      evid[start[j]:end[j]],
                      cmt[start[j]:end[j]],
                      addl[start[j]:end[j]],
                      ss[start[j]:end[j]],
                      {CLPred[j], QPred[j], VCPred[j], VPPred[j], kaPred[j]});

      concentrationPred[start[j]:end[j]]
        = massPred[2, start[j]:end[j]] / VCPred[j];
  }

  cObsPred = lognormal_rng(log(concentrationPred[iObs]), sigma);
}
\end{lstlisting}


It is worth noting that the computational cost of running operations in the \texttt{generated quantities} is relatively small.
While these operations are executed once per iteration, in order to generate posterior samples of the generated quantities, operations in the \texttt{transformed parameters} and \texttt{model} blocks are run and differentiate multiple times per iterations, meaning they amply dominate the computation.
Hence the cost of doing posterior predictive checks, even when it involves solving ODEs, is marginal.
The computational scaling of Stan, notably for ODE-based models, is discussed in the article by Grinsztajn et al (2021) \cite{Grinsztajn:2021}.

{
    For this simple population PK modeling example with a uniform
    study design for all individuals, the PPCs shown in Figures 5 and
    6 are arguably sufficient model diagnostics. In cases where the
    study design and patient populations are more heterogeneous,
    methods that adjust for such heterogeneity are desirable. NPDEs \cite{brendel2006metrics}
    are commonly used in the maximum likelihood context and could be
    applied to point predictions from Bayesian models, e.g., posterior
    mean or median predictions. A similar approach termed probability
    integral transforms (PIT) are used for Bayesian model checking
    \cite{Gelman:2020, Gabry:2019}.

    Standard PPCs that use the same data for model fitting and model
    checking may be over-optimistic particularly when applied to
    highly flexible or overparameterized models. This may be remedied
    by using out-of-sample predictions for PPCs and PITs. In the
    context of population models this means fitting the model to data
    from a subset of individuals and predicting outcomes for the
    remaining individuals. This may be done for an entire data set
    using cross validation. However,
    generating a cross validation predictive
    distribution is computationally expensive and may often be impractical. 
}

\section{Nonlinear pharmacokinetic / pharmacodynamic model}

Now let us consider a PKPD model described in terms of a nonlinear ODE
system that requires the use of a numerical solver.
The patient receives multiple doses at regular time intervals and the drug plasma concentration is recorded over time.

\subsection{Nonlinear ODE model in pharmacokinetics / pharmacodynamics} 
\begin{figure}
    \begin{center}
    \includegraphics[width=5in]{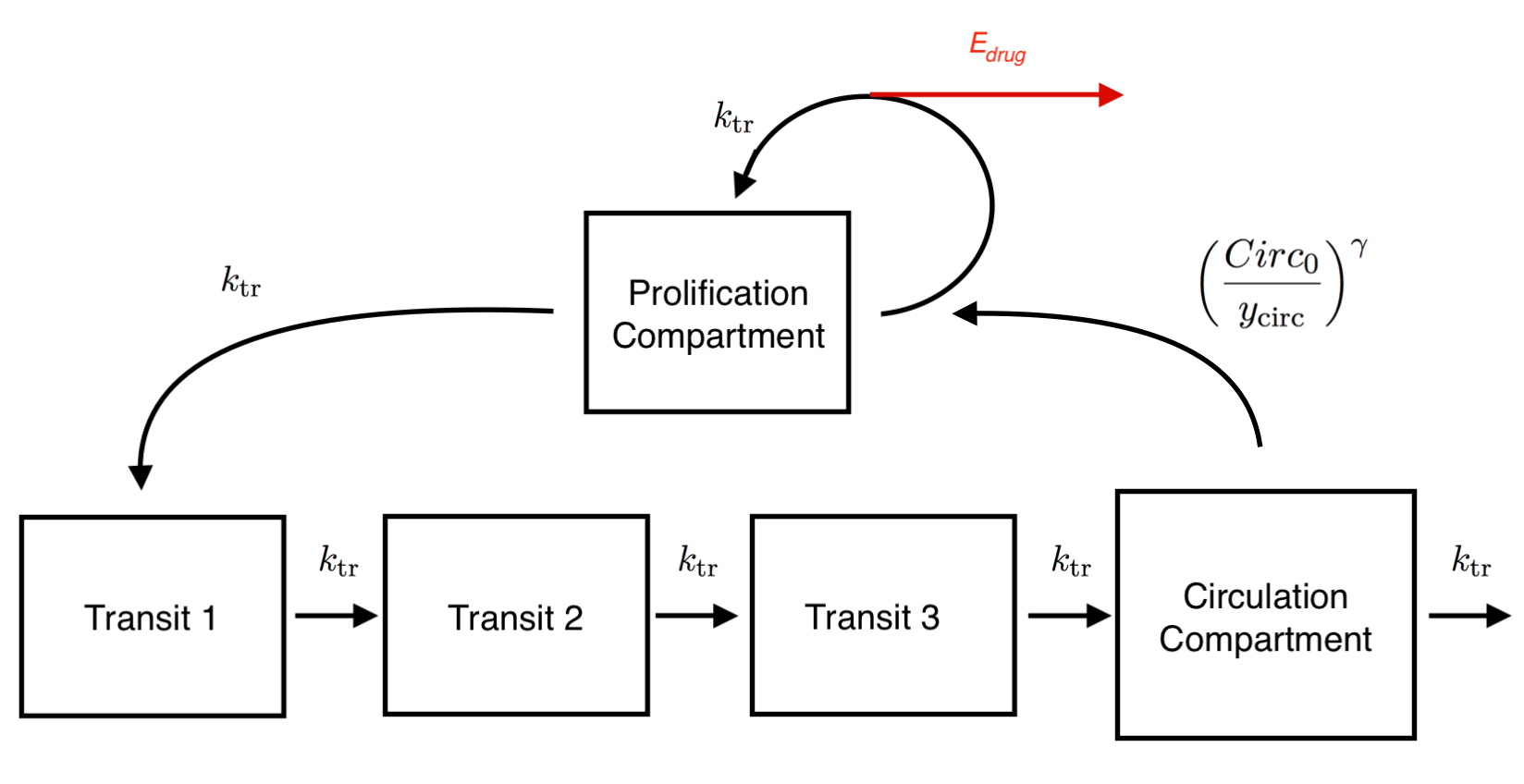}
    \caption{Friberg-Karlsson semi-mechanistic Model.}
    \label{fig:fk_model}
    \end{center}
  \end{figure}

In this the last example, we go back to the single patient
two--compartment model and append it with a PD
model. Specifically, we examine the
Friberg-Karlsson semi-mechanistic model for drug-induced
myelosuppression \cite{3181,2364,2518,3187,3188,3537} with the goal to model the
relation between neutrophil counts and drug exposure.
The model describes a delayed feedback mechanism that keeps the absolute neutrophil count (ANC) at the
baseline ($\text{Circ}_0$) in a circulatory compartment ($y_{\text{circ}}$), as well as the drug
effect that perturbs this meachanism. The delay between
proliferative cells ($y_{\text{prol}}$) and $y_{\text{circ}}$ is modeled by three
transit compartments with mean transit time
\begin{equation}
  \text{MTT} = (3 + 1)/k_{\text{tr}}
\end{equation}
where $k_{\text{tr}}$ is the transit rate constant. 
Figure \ref{fig:fk_model} summarizes the model (see also \cite[Figure 2]{3181}).


The PD likelihood is
\begin{align*}
  \text{ANC} & \sim \text{logNormal}(\log(y_{\text{circ}}), \sigma_{\text{ANC}}),  \\
  y_{\text{circ}}& = f_{\text{FK}}(\text{MTT}, \text{Circ}_{0}, \alpha, \gamma, \hat c),
\end{align*}
where $\hat c=y_{\text{cent}}/V_{\text{cent}}$ is the drug concentration calculated from the PK model, and $f_{\text{FK}}$ solves the non-linear ODE: 
\begin{subequations}\label{eq:FK}
\begin{align}
  \frac{dy_\mathrm{prol}}{dt} &= k_\mathrm{prol} y_\mathrm{prol} (1 - E_\mathrm{drug})\left(\frac{\text{Circ}_0}{y_\mathrm{circ}}\right)^\gamma - k_\mathrm{tr}y_\mathrm{prol}, \\
  \frac{dy_\mathrm{trans1}}{dt} &= k_\mathrm{tr} y_\mathrm{prol} - k_\mathrm{tr} y_\mathrm{trans1}, \\
  \frac{dy_\mathrm{trans2}}{dt} &= k_\mathrm{tr} y_\mathrm{trans1} - k_\mathrm{tr} y_\mathrm{trans2},  \\
  \frac{dy_\mathrm{trans3}}{dt} &= k_\mathrm{tr} y_\mathrm{trans2} - k_\mathrm{tr} y_\mathrm{trans3},  \\
  \frac{dy_\mathrm{circ}}{dt} &= k_\mathrm{tr} y_\mathrm{trans3} - k_\mathrm{tr} y_\mathrm{circ},
\end{align}
\end{subequations}
We use 
\begin{equation*}
E_{\text{drug}} = \mathrm{min}(\alpha \hat c, 1)
\end{equation*}
to model the linear effect of the drug once it has been absorbed in the central compartment.
This effect reduces the proliferation rate and induces a reduction in neutrophil count. 
The upper bound of 1 on $E_\text{drug}$ excludes the scenario where the feedback loop is flipped if $\hat c$ becomes too large.
While we expect that for any reasonable parameter values, $E_\text{drug} < 1$, we should anticipate the possibility that our Markov chains may encounter less well-behaved values as it explores the parameter space.
Encoding such constraints can lead to improved numerical stability when solving the ODE.

We obtain the complete ODE system for the PKPD model by coupling equation \eqref{eq:twocpt} and \eqref{eq:FK}.
Because the equation is nonlinear, we can no longer resort to analytical solutions as we have done in the previous sections.



\subsubsection{Numerically solving ODEs}
To solve an ODE numerically in Stan we first need to define
a function that returns right-hand-side of the ODE, i.e. the derivative of the solution, in the \texttt{functions} block.
The \texttt{functions} block allows users to define functions and is written at the top of the Stan file before the \texttt{data} block.
%
\begin{lstlisting}[style=stan, numbers=none]
functions {
  vector twoCptNeutModelODE(real t, vector y, real[] parms, real[] rdummy, int[] idummy){
    real CL = parms[1];
    real Q = parms[2];
    real V1 = parms[3];
    real V2 = parms[4];
    real ka = parms[5];
    real mtt = parms[6];	
    real circ0 = parms[7];
    real gamma = parms[8];
    real alpha = parms[9];
    real k10 = CL / V1;
    real k12 = Q / V1;
    real k21 = Q / V2;
    real ktr = 4 / mtt;
    vector[8] dydt;
    real conc = y[2] / V1;
    real EDrug = fmin(1.0, alpha * conc);
    real prol = y[4] + circ0;
    real transit1 = y[5] + circ0;
    real transit2 = y[6] + circ0;
    real transit3 = y[7] + circ0;
    real circ = fmax(machine_precision(), y[8] + circ0);

    dydt[1] = -ka * y[1];
    dydt[2] = ka * y[1] - (k10 + k12) * y[2] + k21 * y[3];
    dydt[3] = k12 * y[2] - k21 * y[3];

    // y[4], y[5], y[6], y[7] and y[8] are differences from circ0.
    dydt[4] = ktr * prol * ((1 - EDrug) * ((circ0 / circ)^gamma) - 1);
    dydt[5] = ktr * (prol - transit1);
    dydt[6] = ktr * (transit1 - transit2);
    dydt[7] = ktr * (transit2 - transit3);
    dydt[8] = ktr * (transit3 - circ);

    return dydt;
  }
}
\end{lstlisting}
The above function is an almost direct translation of Eq. \eqref{eq:twocpt} and \eqref{eq:FK}.
The first three components of \texttt{dydt} describes the PK.
The next five components of \texttt{dydt} describe the PD minus the baseline $\mathrm{Circ}_0$.
Writing the ODE as a difference from the baseline means the initial PD conditions is {\bf 0}, as opposed to a parameter dependent value.
This results in better computation, because derivatives of the ODE solution with respect to the initial conditions no longer need to be computed; for more details, see \cite[Section 5.2]{Grinsztajn:2021}.
In addition, we encode a constraint on the circulatory compartment
\begin{equation*}
  y_\text{circ} > \epsilon > 0,
\end{equation*}
where $\epsilon$ is the machine precision and can be interpreted as the smallest non-zero number the computer can handle.
This is to improve numerical stability, especially during the early stages of MCMC exploration when we may need to handle somewhat implausible parameter values.

Stan and Torsten provide several numerical solvers.
In this example we use the Runge-Kutta solver
\texttt{pmx\_solve\_rk45} \cite[Section 3.4]{Torsten:2021}.
The signature of \texttt{pmx\_solve\_rk45} is a bit more sophisticated than that of \texttt{pmx\_solve\_twocpt} and requires the following arguments:
\begin{enumerate}
  \item the name of the user-defined ODE function (\texttt{twoCptNeutModelODE})
  \item the number of states/compartments in the ODE
  \item the event schedule
  \item the bioavailibility fraction, $F$, and the dosing lag time, $t_\mathrm{lag}$ for each compartment (optional)
  \item the tuning parameters for the ODE solver (optional)
\end{enumerate}
Because arguments are nameless in Stan, we can only pass the ODE tuning parameters if we also pass $F$ and $t_\mathrm{lag}$.
By setting $F$ to 1 and $t_\mathrm{lag}$ to 0 for each compartment, we essentially ignore their effect.
This is best done in the \texttt{transformed data} block:
%
\begin{lstlisting}[style=stan, numbers=none]
  int<lower = 1> nCmt = 8;
  real F[nCmt] = rep_array(1.0, nCmt);
  real tLag[nCmt] = rep_array(0.0, nCmt);
\end{lstlisting}
Numerical solvers in Stan and Torsten admit three tuning parameters:
\begin{itemize}
  \item \texttt{rtol}: relative tolerance to determine solution convergence,
  \item \texttt{atol}: absolute tolerance to determine solution convergence,
  \item \texttt{max\_num\_step}: maximum number of steps allowed.
\end{itemize}
Though Stan and Torsten provide
default values, we highly recommend that the user define the ODE
solver control parameters in the \texttt{data} block:
%
\begin{lstlisting}[style=stan, numbers=none]
  real<lower = 0> rtol;
  real<lower = 0> atol;
  int<lower = 0> max_num_step;
\end{lstlisting}
Users should make problem-dependent decisions on \texttt{rtol} and \texttt{atol},
according to the expected scale of the unknowns, so that the error does
not affect our inference. 
For example, when an unknown can be neglected below a
certain threshold without affecting the rest of the dynamic system,
setting \texttt{atol} greater than that threshold avoids spurious and
error-prone computation. For more details, see \cite[Chapter 13]{Stan_users_guide:2021}
and \cite[Section 3.7.5]{Torsten:2021} and references therein.

As before, we solve the ODE within the event schedule in the \texttt{transformed parameters} block:
%
\begin{lstlisting}[style=stan, numbers=none]
transformed parameters{
  vector[nt] cHat;
  vector[nObsPK] cHatObs;
  vector[nt] neutHat;
  vector[nObsPD] neutHatObs;
  matrix[nCmt, nt] y;
  real<lower = 0> parms[9];

  parms = {CL, Q, V1, V2, ka, mtt, circ0, gamma, alpha};

  y = pmx_solve_rk45(twoCptNeutModelODE, nCmt, time, amt, rate, ii, evid, cmt, addl, ss, parms, F, tLag, rtol, atol, max_num_step);

  cHat = y[2, ]' / V1;
  neutHat = y[8, ]' + circ0;

  cHatObs = cHat[iObsPK]; // predictions for observed data records
  neutHatObs = neutHat[iObsPD]; // predictions for observed data records
}
\end{lstlisting}

\subsubsection{Solving PKPD ODEs as a coupled system}

The approach in the last section applies to all models that involve
ODE solutions, but we will not use it here. An acute
observer may have noticed the PKPD model here exhibits a particular
\emph{one-way coupling} structure.
That is, the PK (Eq. \eqref{eq:twocpt})
and PD (Eq. \eqref{eq:FK}) are
coupled through the proliferation cell count
$y_{\text{prol}}$ and $E_{\text{drug}}$, such that the PK 
can be solved independently from the PD. This is what motivates Torsten's coupled solvers
which analytically solves the PK ODEs before
passing the PK solution to the PD ODE. 
The PD ODE is then solved numerically.
Since the dimension of the numerical ODE solution is reduced, in general this coupled strategy is more efficient than
the last section's approach of numerically solving a full ODE system.
To see it in action, let us apply the
coupled solver \texttt{pmx\_solve\_twocpt\_rk45} \cite[Section 3.5]{Torsten:2021} to the same model. We need only make two changes. First, we
modify the ODE function to reflect that only the PD states are to be solved.
\begin{lstlisting}[style=stan, numbers=none]
functions{
  vector twoCptNeutModelODE(real t, vector y, vector y_pk, real[] theta, real[] rdummy, int[] idummy){
    /* PK variables */
    real V1 = theta[3];

    /* PD variable */
    real mtt      = theta[6];
    real circ0    = theta[7];
    real gamma    = theta[8];
    real alpha    = theta[9];
    real ktr      = 4.0 / mtt;
    real prol     = y[1] + circ0;
    real transit1 = y[2] + circ0;
    real transit2 = y[3] + circ0;
    real transit3 = y[4] + circ0;
    real circ     = fmax(machine_precision(), y[5] + circ0);
    real conc     = y_pk[2] / V1;
    real EDrug    = alpha * conc;

    vector[5] dydt;

    dydt[1] = ktr * prol * ((1 - EDrug) * ((circ0 / circ)^gamma) - 1);
    dydt[2] = ktr * (prol - transit1);
    dydt[3] = ktr * (transit1 - transit2);
    dydt[4] = ktr * (transit2 - transit3);
    dydt[5] = ktr * (transit3 - circ);

    return dydt;
  }
}
\end{lstlisting}

Note that  we pass in PD and PK states as separate arguments, $y$
and $y_{\text{PK}}$ respectively.
The above function only returns $\mathrm d y/ \mathrm d t$, while
$y_{\text{PK}}$ is solved internally using an analytical solution,
meaning users do not need to explicitly call \texttt{pmx\_solve\_twocpt}.

Then we replace \texttt{pmx\_solve\_rk45} with
\texttt{pmx\_solve\_twocpt\_rk45} call.
\begin{lstlisting}[style=stan, numbers=none]
  x = pmx_solve_twocpt_rk45(twoCptNeutModelODE, nOde, time, amt, rate, ii, evid, cmt, addl, ss, parms, biovar, tlag, rtol, atol, max_num_step);
\end{lstlisting}

\subsection{Building the remaining coding blocks}

We omit the \texttt{data} block, but note that it is similar to the one we constructed in previous sections.
The key difference is we now include measurements for the absolute neutrophil count.
The \texttt{parameters} block now contains the PD variables:
%
\begin{lstlisting}[style=stan, numbers=none]
parameters{
  real<lower = 0> CL;
  real<lower = 0> Q;
  real<lower = 0> V1;
  real<lower = 0> V2;
  //  real<lower = 0> ka; // ka unconstrained
  real<lower = (CL / V1 + Q / V1 + Q / V2 +
		sqrt((CL / V1 + Q / V1 + Q / V2)^2 -
		     4 * CL / V1 * Q / V2)) / 2> ka; // ka > lambda_1
  real<lower = 0> mtt;
  real<lower = 0> circ0;
  real<lower = 0> alpha;
  real<lower = 0> gamma;
  real<lower = 0> sigma;
  real<lower = 0> sigmaNeut;
}
\end{lstlisting}

The \texttt{model} block is similar to that in Section \ref{sec:twocpt}:
\begin{lstlisting}[style=stan, numbers=none]
model{
  CL ~ lognormal(log(CLPrior), CLPriorCV);
  Q ~ lognormal(log(QPrior), QPriorCV);
  V1 ~ lognormal(log(V1Prior), V1PriorCV);
  V2 ~ lognormal(log(V2Prior), V2PriorCV);
  ka ~ lognormal(log(kaPrior), kaPriorCV);
  sigma ~ normal(0, 1);

  mtt ~ lognormal(log(mttPrior), mttPriorCV);
  circ0 ~ lognormal(log(circ0Prior), circ0PriorCV);
  alpha ~ lognormal(log(alphaPrior), alphaPriorCV);
  gamma ~ lognormal(log(gammaPrior), gammaPriorCV);
  sigmaNeut ~ normal(0, 1);

  cObs ~ lognormal(log(cHatObs), sigma); // observed data likelihood
  neutObs ~ lognormal(log(neutHatObs), sigmaNeut);
}
\end{lstlisting}

\subsubsection{Posterior predictive checks}
We hope by now reader has developed the habit of performing
PPC on every model. Since we have both PK (drug
concentration) and PD (neutrophil count) observations, the PPC should
be conducted on both.
\begin{lstlisting}[style=stan, numbers=none]
generated quantities{
  real cObsPred[nt];
  real neutObsPred[nt];

  for(i in 1:nt){
    if(time[i] == 0){
      cObsPred[i] = 0;
    }else{
      cObsPred[i] = lognormal_rng(log(cHat[i]), sigma);
    }
    neutObsPred[i] = lognormal_rng(log(neutHat[i]), sigmaNeut);
  }
}
\end{lstlisting}

It is possible to only run the generated quantities block based on a fitted model using \texttt{cmdstanr}'s \texttt{generate\_quantities} routine.
This is useful when we change the generated quantities, but not the rest of a model we have already fitted.
The compiled model and the fit are respectively stored in the \texttt{mod} and \texttt{fit} objects in R.
We then run:
\begin{lstlisting}[language=R, numbers=none]
fit.gq <- mod$generate_quantities(fit)
\end{lstlisting}
and use the results for PPC (Figure \ref{fig:neutro_ppc}).


 \begin{figure}
  \includegraphics[width=\textwidth]{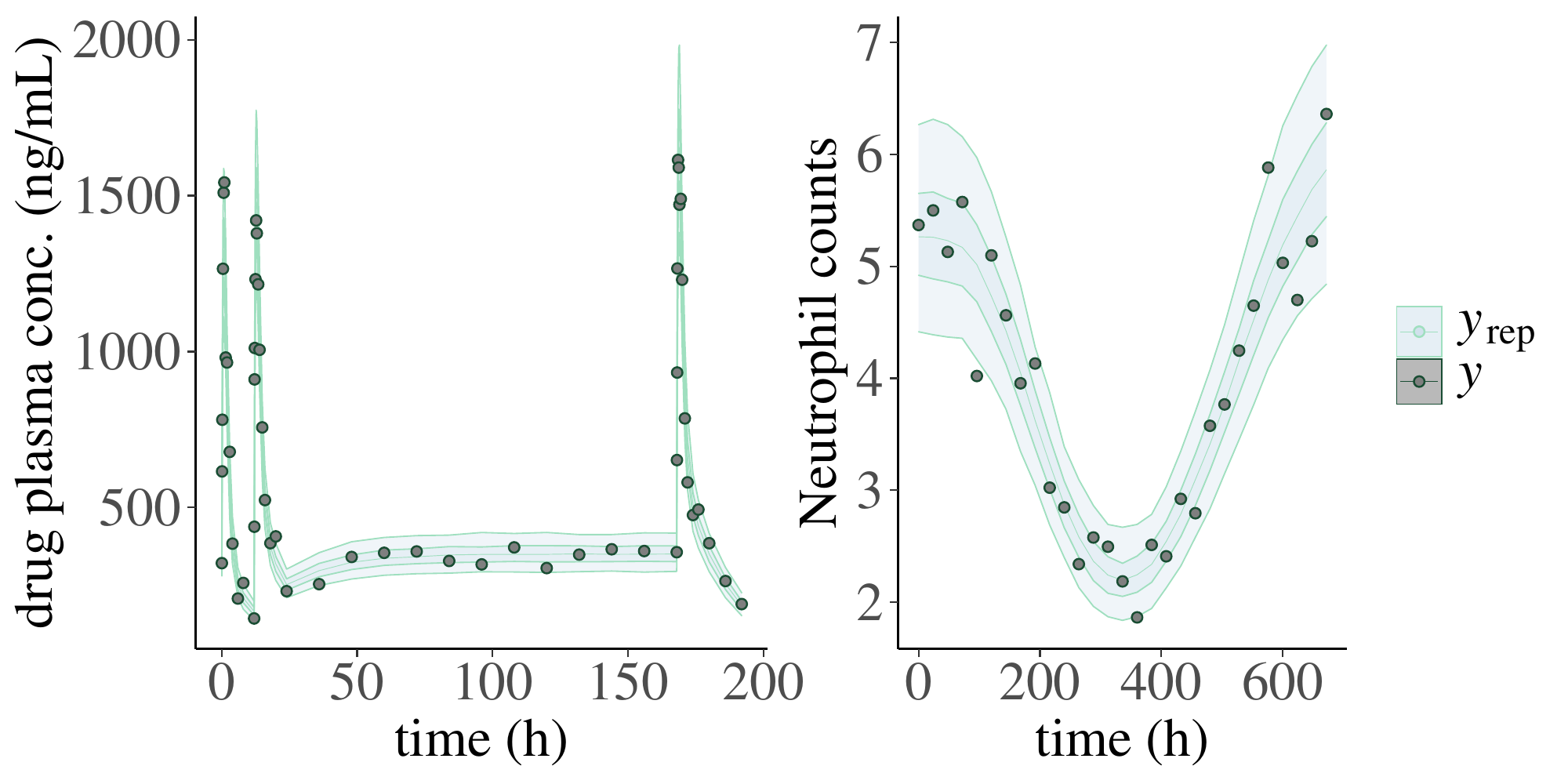}
   \caption{Posterior predictive checks for the PKPD model}
 \end{figure}

\section{Discussion}

Stan provides an expressive language to build models, state-of-the-art algorithms to fit these models,
and a host of easy-to-use diagnostics.  
Torsten complements Stan with a suite of routines which solve ODEs within the context of a clinical event schedules.
Together, Stan and Torsten are potent tools when working through the tangled steps of a Bayesian workflow for PKPD modeling.

\subsection{Current and potential role for Stan and Torsten for pharmacometrics applications} \label{sec:pmx_role}
We can apply Stan/Torsten to a large palette of generative models, both for inference and simulation. 
Applications range from simple linear regression to complex multi-scale Quantitative Systems Pharmacology
models. Compared to specialized pharmacometrics tools such as NONMEM\textsuperscript{\textregistered},
Stan/Torsten is particularly well--suited for cases where more flexibility is desired. 
This includes models with

\begin{itemize}
\item random effects distributions other than normal,
\item prior distributions other than the limited set
  available in existing pharmacometrics tools,
\item multiple submodels with different random effect structures.
\end{itemize}

It's important to recognize that MCMC, including the HMC scheme used by Stan/Torsten,
can be computationally intensive, notably when fitting hierarchical models which require us
to numerically solve ODEs.
This can be especially frustrating during the initial model
exploration stage of a project. For such exploratory analyses access
to a rapid approximate Bayesian inference engine may be
desirable. Stan/Torsten includes two optimization-based inference
engines, one for estimation of posterior modes and one for variational
inference. 
These algorithms attempt to
simultaneously optimize over the entire joint posterior distribution
of all model parameters. This process can be relatively slow and error
prone when trying to optimize over the large number of population and
individual--level parameters of a typical population pharmacometrics model.
This contrasts with typical mixed effects modeling programs that use
algorithms specialized for a more limited range of models---usually
employing an alternating sequence of lower dimensional optimization
problems. 

For applications that may be implemented with typical pharmacometrics tools, the
choice between those and Stan/Torsten comes down to the trade-offs
between flexibility, doing accurate Bayesian inference and computation time.

We would also like to point out that Stan is not the only probabilistic programing language that is actively under development.
PyMC3 \cite{salvatier_probabilistic_2016}, TensorFlow Probability \cite{Dillon:2017, Lao:2020}, and Turing \cite{ge_turing_2018},
among others, provide similar modeling capabilities.
A full review and comparison of these languages is however beyond the scope of this paper.

\subsection{Preview of part 2} \label{sec:part2}
In part 2 of this tutorial, we plan to build on the material we have covered thus far and tackle more advanced topics, including:
\begin{itemize}
  \item \textit{Improving the performance of HMC}, using within-chain parallelization for population models and Torsten's dedicated group solvers.
  \item \textit{Advanced diagnostic tools}, namely divergent transitions which can flag bias in our posterior samples. Stan makes these diagnostics readily available. 
  \item \textit{Fake data simulation and analysis,} in particular prior predictive checks as a way to understand and build priors, fitting the model to fake data as an imperfect tool to troubleshoot Bayesian inference, and an overview of the more sophisticated but computationally demanding simulation-based calibration \cite{Talts:2020}.
  \item \textit{Performance tuning of ODE models}, such as solver selection, accurarcy control, as well as stability issues.
\end{itemize}
We will dive into these subjects by examining more advanced models and utilizing techniques
such as reparameterization, within-chain parallelization, and pooling
multiple data sources. 
We will also discuss ongoing developments with Stan and Torsten, such as tools to handle larger scale ODEs and plans to leverage parallelization.

\bibliographystyle{vancouver}

\clearpage

\end{document}